\documentclass[aps,prb,superscriptaddress,twocolumn]{revtex4-2}

\usepackage{graphicx} 
\usepackage{hyperref}
\usepackage{amsmath} 
\usepackage{amssymb} 
\usepackage{amsthm}
\usepackage{amsfonts}
\usepackage{color}
\usepackage{bm}
\usepackage{bbm}
\usepackage{url}
\usepackage{booktabs}
\usepackage{hyperref}
\hypersetup{pdfpagemode=FullScreen,colorlinks=true,breaklinks,urlcolor=blue,linkcolor=blue,citecolor=blue}

\begin{document}

\global\long\def\id{\mathbbm{1}}
\global\long\def\ui{\mathbbm{i}}
\global\long\def\ud{\mathrm{d}}

\title{Probing critical phases in quasiperiodic systems via subsystem information capacity}

\author{Huaijin Dong}
\affiliation{School of Physics and Institute for Quantum Science and Engineering, Huazhong University of Science and Technology, Wuhan 430074, China}
\author{Long Zhang}
\email{lzhangphys@hust.edu.cn}
\affiliation{School of Physics and Institute for Quantum Science and Engineering, Huazhong University of Science and Technology, Wuhan 430074, China}
\affiliation{Hefei National Laboratory, Hefei 230088, China}


\begin{abstract}

We systematically investigate the entanglement and information dynamics of quasiperiodic systems across their extended, critical, and localized phases, aiming to identify dynamical signatures that can reveal the multifractal spatial structure of critical states and distinguish critical phases from the extended and localized regimes. Focusing on the generalized Aubry-Andr\'e-Harper model, we complement the half-chain entanglement entropy with the spatially resolved subsystem information capacity (SIC) and demonstrate that critical states exhibit pronounced spatial heterogeneity absent in the extended and localized phases. In the steady state, the SIC reveals a stepwise ramp as a function of subsystem size, reflecting an underlying fragmentation of the chain into weakly connected subregions. Dynamically, information initially localized within such a subregion can undergo coherent long-lived oscillations, dubbed {\it subregion echoes}, whose period scales with the subregion length, in quantitative agreement with a quasiparticle picture of confined quasiparticle reflections. We trace this internal fragmentation to the incommensurately distributed zeros (IDZs) in the off-diagonal hopping terms of the Hamiltonian. To establish the generality of the SIC as a diagnostic tool, we further apply it to a mobility-edge phase with coexisting extended and localized states and to a critical phase that does not originate from IDZ fragmentation, and show that the SIC can cleanly distinguish these scenarios through their distinct steady-state profiles, initial-site sensitivities, and the presence or absence of subregion echoes. Our results {demonstrate that the SIC serves} as a powerful real-space probe for diagnosing critical phases and uncovering the bottlenecked connectivity that underlies the multifractal structure of critical states.

\end{abstract}

\maketitle

\section{Introduction}

Anderson localization, arising from destructive interference among multiple scattering paths in a disordered medium, 
is a cornerstone phenomenon in the study of quantum transport~\cite{Anderson1958,Evers2008}. 
Situated between perfectly periodic and fully random systems, quasiperiodic lattices
provide a distinct and highly tunable setting for localization physics~\cite{Roati2008,Lucioni2011,Luschen2018,Kohlert2019,An2021,WangY2022,Gao2025,Xiao2021,Li2023,Shimasaki2024,Huang2025,Zhu2026}.
The prototypical quasiperiodic Aubry-Andr\'e (AA) model exhibits a direct transition between fully extended and fully localized spectra~\cite{Aubry1980}, 
whereas more generalized quasiperiodic models can host single-particle mobility edges separating extended and localized states within the same spectrum~\cite{Luschen2018,Kohlert2019,An2021,WangY2022,Gao2025,Soukoulis1982,DasSarma1988,Boers2007,Biddle2010,Ganeshan2015,Gopalakrishnan2017,Li2017,Li2020,Wang2020b,Roy2021,Wang2023,Hu2025}. 
Most intriguingly, quasiperiodic systems may also host multifractal critical states and even critical phases that interpolate between the fully extended and localized regimes~\cite{Xiao2021,Li2023,Shimasaki2024,Huang2025,Zhu2026,Wang2020a,Liu2015,Deng2019,RoyN2021a,Wang2022,Liu2022,Zhou2023,Gonifmmode2023,Roy2023,Lin2023,Goncalves2024,Yang2024,Duncan2024,Zhou2026}. 
Such critical states are neither Bloch-like nor exponentially localized; instead, they are characterized by multifractal wave functions, scale-invariant spatial structures, and anomalous transport dynamics~\cite{Wang2020a}.
In some generalized quasiperiodic models, such criticality can occupy a finite region of parameter space, forming a genuine critical phase rather than merely an isolated transition point~\cite{Wang2020a,Liu2015,Deng2019,RoyN2021a,Wang2022,Liu2022,Zhou2023,Gonifmmode2023,Roy2023,Lin2023,Goncalves2024,Yang2024,Duncan2024,Zhou2026}.
Recent theoretical advances have further sharpened the characterization of critical states and the associated phase boundaries, 
establishing a rigorous framework for identifying criticality and the mobility edges separating critical, extended, and localized regimes~\cite{Zhou2023,Zhou2026}.

Despite these conceptual advances, a detailed understanding of the internal spatial architecture of critical phases---namely, how quantum information propagates through a multifractal landscape---remains incomplete.
Conventional diagnostics, such as the fractal dimension extracted from eigenstate inverse participation ratios, provide primarily static and globally averaged characterizations of the spectrum~\cite{Wang2020a,Deng2019,RoyN2021a,Wang2022}. Dynamical probes such as wave-packet expansion can distinguish extended, critical, and localized phases through their distinct transport exponents~\cite{Huang2025,Wang2020a}.
However, these probes generally average over the internal spatial organization of critical states and therefore do not directly resolve the heterogeneous connectivity inherent to multifractal wave functions. 
This limitation motivates the search for spatially resolved dynamical probes capable of characterizing local quantum-information flow rather than global transport alone.
Entanglement and information dynamics provide a natural framework in this direction, since the spreading of entanglement and quantum information is closely tied to the structure of eigenstates and the transport properties of the system~\cite{Bardarson2012,Serbyn2013,Nanduri2014,Deng2017,Fan2017,Zhao2020,Ghosh2021,RoyN2021b,Zhang2022}. While the basic patterns of global entanglement growth in extended and localized phases are well understood---ballistic spreading in the extended phase~\cite{Calabrese2005} and strong suppression in the localized phase~\cite{Bardarson2012,RoyN2021b}---a systematic spatially resolved dynamical characterization of critical phases is far from complete. 
Recent studies have revealed that the entanglement entropy (EE) in critical phases exhibits a logarithmic scaling with subsystem size~\cite{Gonifmmode2024}.
Nevertheless, how the multifractal nature of critical wave functions manifests in real‑time entanglement dynamics, and whether critical phases of different microscopic origins can be distinguished by their real‑space information-flow patterns, remain largely unexplored. Motivated by these questions, the recently introduced subsystem information capacity (SIC) offers a particularly appealing real-space diagnostic~\cite{Chen2025,Qing2026}. { Unlike static measures such as the inverse participation ratio (IPR), which provide only a global average over eigenstate spatial distributions, the SIC tracks} how quantum information initially encoded at a single site diffuses through the system, providing a spatially resolved and site-selective probe of information transport that is ideally suited to capture the heterogeneous connectivity underpinning critical phases.

In this work, we systematically investigate the entanglement and information dynamics of the generalized Aubry-Andr\'e-Harper (GAAH) model across its extended, critical, and localized phases, focusing on how the multifractal spatial structure of critical states manifests itself in real-space information dynamics. 
We combine the conventional half-chain EE, which characterizes global entanglement dynamics, with the SIC, which provides a spatially resolved probe of information dynamics.
We demonstrate that, while the half-chain EE can separate localized from delocalized behavior, it fails to resolve the rich spatial structure of critical states. 
The SIC, by contrast, reveals pronounced spatial heterogeneity characteristic of critical states, which we trace back to the incommensurately distributed zeros (IDZs) in the off-diagonal hopping terms. These IDZs partition the chain into nearly disconnected subregions, leading to a stepwise ramp in the steady-state SIC and to coherent long-lived subregion echoes in its dynamics. We further show that these oscillatory echoes are quantitatively explained by a quasiparticle picture in which quasiparticles are confined within the IDZ-delimited subregions, undergoing repeated reflections. To establish the generality of the SIC as a diagnostic tool, we extend our analysis to two additional settings: a phase with a single-particle mobility edge (SPME) separating extended and localized states and a critical phase that does not originate from IDZ fragmentation, and demonstrate that the SIC can cleanly distinguish these scenarios through their distinct steady-state profiles, initial-site sensitivities and the appearance of subregion echoes. Our results {demonstrate that the SIC provides} a powerful and readily measurable probe for distinguishing critical phases and uncovering the bottlenecked connectivity that underlies the multifractal architecture of critical states.

The remainder of this paper is organized as follows. Section~\ref{Sec: Model} introduces the GAAH model and its phase diagram. Section~\ref{Sec: Entanglement Dynamics} presents the half-chain EE dynamics. Section~\ref{Sec: SIC}  is devoted to the SIC, where we discuss the steady-state spatial fingerprints (Sec.~\ref{Sec: Steady-state SIC}) and the subregion echoes arising from IDZ-induced fragmentation with a quasiparticle interpretation (Sec.~\ref{Sec: SIC Dynamics}). In Sec.~\ref{Sec: Comparison}, we compare the SIC signatures of the critical phase originating from IDZ fragmentation with those of two other nontrivial localization scenarios: an SPME phase (Sec.~\ref{Sec: SPME phase}) and a critical phase without IDZs (Sec.~\ref{Sec: non-IDZ critical phase}), and summarize their distinguishing features (Sec.~\ref{Sec: Summary}). We conclude with a summary and outlook in Sec.~\ref{Sec: Conclusion}.
{Additional details are provided in the Appendixes.}


\section{Generalized Aubry-Andr\'e-Harper model}\label{Sec: Model}

We consider the 1D GAAH model~\cite{Hatsugai1990,Han1994}, whose Hamiltonian is given by
\begin{equation}
    H_{\rm GAAH} = \sum_{j}\left[ \left(t_j  c_j^\dagger c_{j+1}+{\rm H.c.}\right) +V \cos(2\pi \beta j + \phi)c_j^\dagger c_j\right],    \label{eq1}
\end{equation}
where the nearest-neighbor hopping strength $t_j=1 + \mu \cos \left[ 2\pi \beta \left( j + \frac{1}{2} \right) + \phi \right]$, and $c_j$ ($c^\dagger_j$) denotes the fermionic annihilation (creation) operator at site $j$. 
The parameter $\mu$ controls the amplitude of the off-diagonal modulation,
while $V$ represents the strength of the incommensurate on-site potential. 
{Note that although $t_j$ can change sign for $\mu>1$, such sign changes can be removed by a local gauge transformation; 
the essential ingredient is the zero or near-zero magnitudes of $|t_j|$, which define the IDZs discussed below.}
The incommensurate wave vector is chosen as the inverse golden ratio, $\beta = (\sqrt{5}-1)/2$.
For a finite system, it is approximated by $\beta=\lim_{n \rightarrow \infty} (F_{n-1} / F_n)$, 
where the Fibonacci sequence is defined by $F_n = F_{n-1} + F_{n-2}$ with $F_0=F_1=1$.
To impose periodic boundary conditions consistently, we set the system length to $L = F_n$ and take
$\beta = F_{n-1} / F_{n}$. Throughout this work, we fix $\phi = 0$.

\begin{figure}
    \centering
    \includegraphics[width=0.99\linewidth]{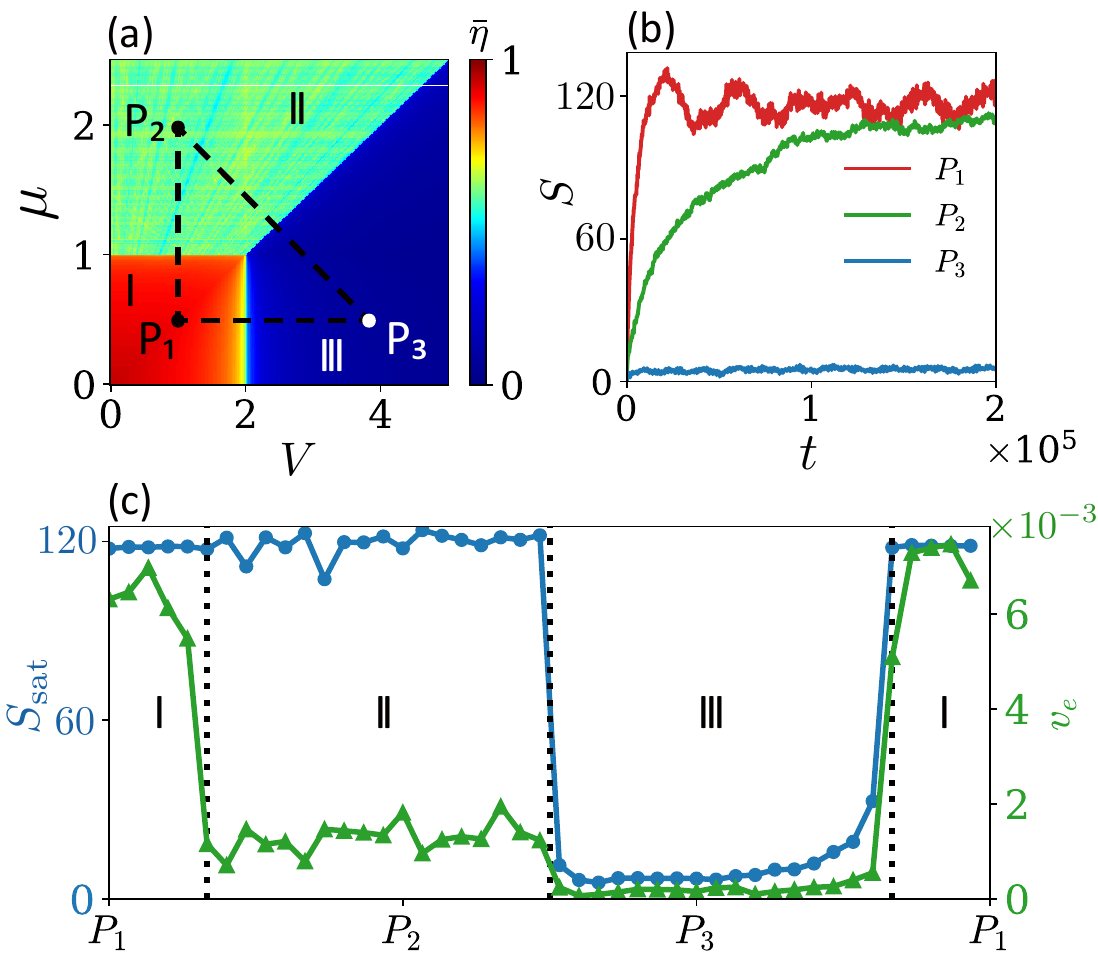}
    \caption{(a) Phase diagram of the GAAH model, characterized by mean fractal dimension $\bar{\eta}$ as a function of $V$ and $\mu$, {estimated for finite} system size $L = F_{14} = 610$. Regions I, II, and III correspond to the extended, critical, and localized phases, respectively. (b) Time evolution of the half‑chain EE for representative points in the three phases. The parameters $(V, \mu)$ are taken from the points labeled $P_1\,(1.0, 0.5)$, $P_2\,(1.0, 2.0)$, and $P_3\,(4.0, 0.5)$ in panel (a). (c) Saturation value $S_{\rm sat}$ (blue curve with dots) and early‑time growth velocity $v_e$ (green curve with triangles) as the parameters $(V, \mu)$ are varied clockwise along the piecewise path indicated by the dashed line in panel (a): $V = 1.0$, $\mu = 2.5 - 0.5V$, and $\mu = 0.5$, starting from point $P_1$. The dotted vertical lines mark the phase boundaries.
    {Here $S_\mathrm{sat}$ is obtained by averaging over the time window $t \in [2 \times 10^6, 4 \times 10^6]$, and $v_e$ is extracted from a linear fit to the initial growth over $t \in [0, 1 \times 10^4]$.}
    }\label{fig1}
\end{figure}

The phase diagram is determined by analyzing the fractal dimension. 
For the $m$th eigenstate $|\psi_m\rangle =\sum_{j=1}^{L} u_{m, j} c_j^\dagger |{\rm vac}\rangle$,
the IPR is defined as $\mathrm{IPR}(m) = \sum_{j} |u_{m, j}|^4$.
The corresponding fractal dimension is then obtained from the scaling $\eta = - \lim_{L \rightarrow \infty} \ln \mathrm{IPR} / \ln L$. 
In the thermodynamic limit, $\eta \rightarrow1$ ($\eta \rightarrow 0$) indicates an extended (localized) state, and an intermediate value $0 < \eta < 1$  characterizes a critical state.
Since the GAAH model hosts pure phases with no mobility edges~\cite{Liu2015}, 
the global phase behavior can be efficiently captured by the mean fractal dimension~\cite{Wang2020a}, defined as
 \begin{align}
 \bar{\eta} = - \lim_{L \rightarrow \infty} \frac{\ln \mathrm{MIPR}}{\ln L},
 \end{align}
 where $\mathrm{MIPR} = L^{-1} \sum_{m = 1}^{L} \mathrm{IPR}(m)$ denotes the average inverse participation ratio over all eigenstates.
 Figure \ref{fig1}(a) displays the resulting phase diagram as a function of $\mu$ and $V$.
 Region I corresponds to the extended phase ($\bar{\eta} \rightarrow 1$), region II to the critical phase ($0 < \bar{\eta} < 1$), 
 and region III to the localized phase ($\bar{\eta} \rightarrow 0$). 
In the limit $\mu = 0$, the model reduces to the well-known AA model~\cite{Aubry1980}, for which the extended-localized transition occurs at $V=2$. 
The Hamiltonian in Eq.~\eqref{eq1} exhibits a duality between $\mu$ and $V$ under the exchange $\mu \leftrightarrow V/2$~\cite{Ino2006}.
The self-dual line $V = 2\mu$ corresponds to the critical-to-localized transition, 
while the extended-to-critical transition at $\mu = 1$ is determined by 
the pure off-diagonal modulation limit, consistent with the phase diagram 
shown in Fig.~\ref{fig1}(a).


\section{Global entanglement dynamics}\label{Sec: Entanglement Dynamics}

Following a quantum quench from an unentangled product state, we examine the EE dynamics of the three distinct phases under unitary evolution $| \Psi(t) \rangle = e^{-\ui H_{\rm GAAH}t} | \Psi(0) \rangle$, where $H_{\rm GAAH}$ is the GAAH Hamiltonian specified in Eq.~\eqref{eq1}.
The initial state is chosen as a bipartite configuration $| \Psi(0) \rangle=| 11\dots 100\dots 0 \rangle$, with the left half of the chain fully occupied and the right half empty. 
The half‑chain EE is defined as
\begin{align}
S (t) = - \operatorname{Tr} [\rho_{L/2} (t) \ln \rho_{L/2} (t)],
\end{align}
where $\rho_{L/2} (t)$ denotes the reduced density matrix of the left half of the system.
{Since the considered model is non-interacting, all calculations are performed using single-particle correlation matrix formalism~\cite{Peschel2003,Peschel2009}.}

Figure \ref{fig1}(b) displays the time evolution of the half‑chain EE for representative parameter sets $(V, \mu)$ corresponding to the three distinct phases identified in Fig.~\ref{fig1}(a):
point $P_1\, (1.0,0.5)$ (extended), point $P_2\, (1.0,2.0)$ (critical), and point $P_3\, (4.0, 0.5)$ (localized).
In the extended phase, the EE grows rapidly and subsequently oscillates around a well‑defined saturation value.
In the localized phase, the EE saturates at an extremely low value, reflecting strong suppression of entanglement propagation. 
In the critical phase, the EE eventually approaches a saturation value comparable to that of the extended phase, 
but the growth is markedly slower, indicating that although entanglement ultimately spreads across the entire system, its propagation is significantly hindered.

To quantitatively distinguish the EE dynamics of different phases, we employ two complementary quantities: 
the saturation value $S_{\rm sat}$ and the early‑time growth velocity $v_e$~\cite{Qing2026}.
The saturation value $S_{\rm sat}$ is obtained by averaging the EE over a long time interval after it has reached a steady regime; 
it reflects the system's overall capacity to sustain entanglement. 
The early‑time velocity $v_e$ is extracted from a linear fit to the initial growth of the EE and characterizes how rapidly entanglement spreads through the system.
Figure~\ref{fig1}(c) shows $S_{\rm sat}$ and $v_e$ as functions of the parameters $(V, \mu)$ varied clockwise along the piecewise path defined by
$V = 1.0$, $\mu=2.5-0.5V$ and $\mu = 0.5$ in Fig.~\ref{fig1}(a), starting from point $P_1$.
Regarding the saturation value $S_{\rm sat}$, the extended and critical phases exhibit similar magnitudes, 
both substantially larger than that of the localized phase. 
In contrast, the early‑time growth velocity $v_e$ is vastly larger in the extended phase than in the other two phases.
A closer comparison reveals that the critical phase possesses a slightly higher $v_e$ than the localized phase.
These findings suggest that, from a global perspective, the critical phase resembles the extended phase in allowing entanglement to permeate the entire chain, 
yet from a microscopic standpoint it behaves more like the localized phase, with pronounced hindrances that slow the initial spread of entanglement.

However, the half‑chain EE provides only a coarse‑grained view of the distinct phases in a quasiperiodic system, and it is particularly limited in characterizing the critical phase. A complete understanding of the critical phase requires a spatially resolved probe that can track how quantum information navigates the underlying multifractal landscape.

\section{Subsystem Information Capacity}\label{Sec: SIC}

In this section, we employ the SIC, introduced in Ref.~\cite{Chen2025}, 
to probe information dynamics from a local, spatially resolved perspective.
In contrast to the global view provided by the half‑chain EE, the SIC tracks how quantum information, initially localized at a single site, 
gradually diffuses throughout the system or remains confined to a local region as a function of the subsystem size.
This fine‑grained analysis { exposes} the underlying multifractal structure of critical states.

\begin{figure}
    \centering
    \includegraphics[width=0.95\linewidth]{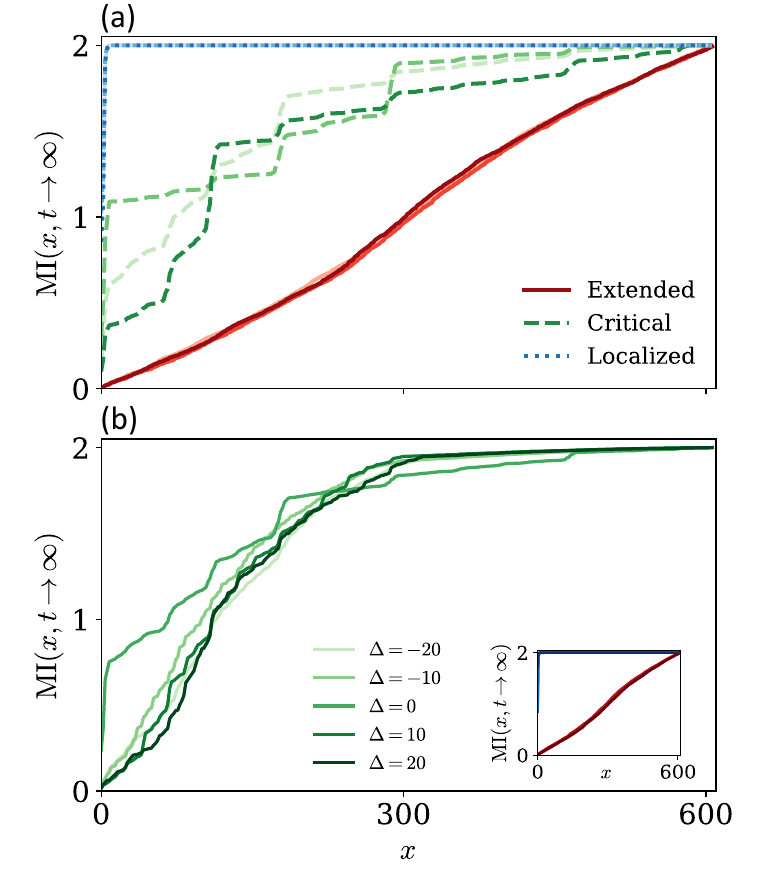}
    \caption{(a) Steady‑state SIC as a function of the subsystem length $x$ following a quantum quench to different phases. The system size is $L = 610$. The red solid lines correspond to the extended phase with parameters $(V, \mu) = (1.0, 0.5)$, $(1.0, 0.6)$, and $(1.0, 0.7)$; the green dashed lines correspond to the critical phase with $(V, \mu) = (1.0, 2.0)$, $(1.2, 1.9)$, and $(1.4, 1.8)$; the blue dotted lines correspond to the localized phase with $(V, \mu) = (4.0, 0.5)$, $(3.8, 0.5)$, and $(3.6, 0.5)$. In each family of curves, the color varies from light to dark in the order listed. (b) Steady‑state SIC in the critical phase [$(V, \mu) = (1.0, 2.0)$] for different spatial offsets $\Delta$ of the initially entangled site $E = L/2 + \Delta$. 
 Inset: Steady‑state SIC in the extended {phase [$(V, \mu) = (1.0, 0.5)]$ and the localized phase [$(V, \mu) = (4.0, 0.5)$]} for the same set of offsets $\Delta$.} \label{fig2}
\end{figure}

\subsection{Steady-state SIC}\label{Sec: Steady-state SIC}

As in the above analysis of the entanglement growth, we consider the same quantum quench protocol starting from the bipartite product state $| 11\dots 100\dots 0 \rangle$. 
To probe local information dynamics, we introduce a reference qubit $R$ that is maximally entangled with a single site $E$ of the chain, resulting in a Bell pair
\begin{equation}
    |\Phi\rangle_{E,R} = \frac{1}{\sqrt{2}} (| 0 \rangle_E | 1 \rangle_R -\ui | 1 \rangle_E | 0 \rangle_R),
\end{equation}
while all other sites remain in an unentangled product state.
{For the case where the encoding site $E$ is at the center of the chain, the total initial state takes the form $| 11\dots 1 \rangle_{\rm Left} \otimes | \Phi \rangle_{E,R} \otimes | 00\dots 0 \rangle_{\rm Right}$. After site $E$ evolves under the GAAH Hamiltonian together with the rest of the chain for a time $t$}, the SIC between the reference qubit $R$ and a subsystem $A$ of length $x=|A|$ centered on site $E$ is defined as the mutual information
\begin{equation}
    {\rm MI}(x, t) = S_A (t) + S_R (t) - S_{AR}(t).
\end{equation}
Here, $S_\alpha$ denotes the von Neumann entropy of $\alpha$; for example, the entropy of subsystem $A$ is $S_A (t) = - \operatorname{Tr} [\rho_A (t) \log \rho_A (t)]$, 
with $\rho_A (t)$ the reduced density matrix of $A$ at time $t$.
We adopt the base‑2 logarithm so that the mutual information takes values in the interval ${\rm MI} \in [0, 2]$. 
The upper bound ${\rm MI}=2$ indicates that all of the quantum information initially encoded at site $E$ can be fully recovered from subsystem $A$.

We first focus on the steady‑state SIC, ${\rm MI}(x, t \rightarrow \infty)$, 
as a function of the subsystem length $x$,
which captures the long‑time spatial distribution of the initially localized information. 
As shown in Fig.~\ref{fig2}(a), the steady‑state SIC profile exhibits a linear ramp in the extended phase, {indicating ballistic information transport mediated by extended states}~\cite{Calabrese2005}, and a step‑like shape in the localized phase, signaling strong information trapping. These observations are consistent with the findings reported in Ref.~\cite{Qing2026}. 
More importantly, our results further demonstrate that the critical phase displays a distinct intermediate behavior: the SIC still attains its maximum value at $x=L$, confirming that information eventually permeates the entire system, yet its growth with $x$ proceeds via a ramp punctuated by multiple steps, in contrast to the smooth increase observed in the extended phase. This {\it stepwise} ramp reflects the underlying multifractal structure of critical states, 
which imposes spatially sensitive hindrances on the local information propagation.

To further expose this underlying structure, we vary the location of the initially entangled site $E$ by setting $E=L/2+\Delta$, 
where $\Delta$ denotes a spatial offset from the chain center. 
We then examine how the steady‑state SIC depends on this initial position of the entangled qubit. 
The results are presented in Fig.~\ref{fig2}(b). 
In the extended and localized phases, introducing a finite offset $\Delta$ produces no discernible change in the steady‑state SIC profile, as illustrated in the inset. 
In striking contrast, for the critical phase, different values of $\Delta$ induce substantial alterations in the shape of the SIC curve. 
Notably, for certain offsets, the stepwise ramp characteristic of the critical phase can be largely smeared out or even erased, underscoring 
the sensitivity of the SIC profile to the exact location of the initially encoded information. 
This behavior directly indicates that the interior of the critical system is far from homogeneous, unlike the spatially uniform extended and localized phases.

\subsection{Subregion echoes in SIC dynamics}\label{Sec: SIC Dynamics}

\begin{figure}
    \centering
    \includegraphics[width=0.98\linewidth]{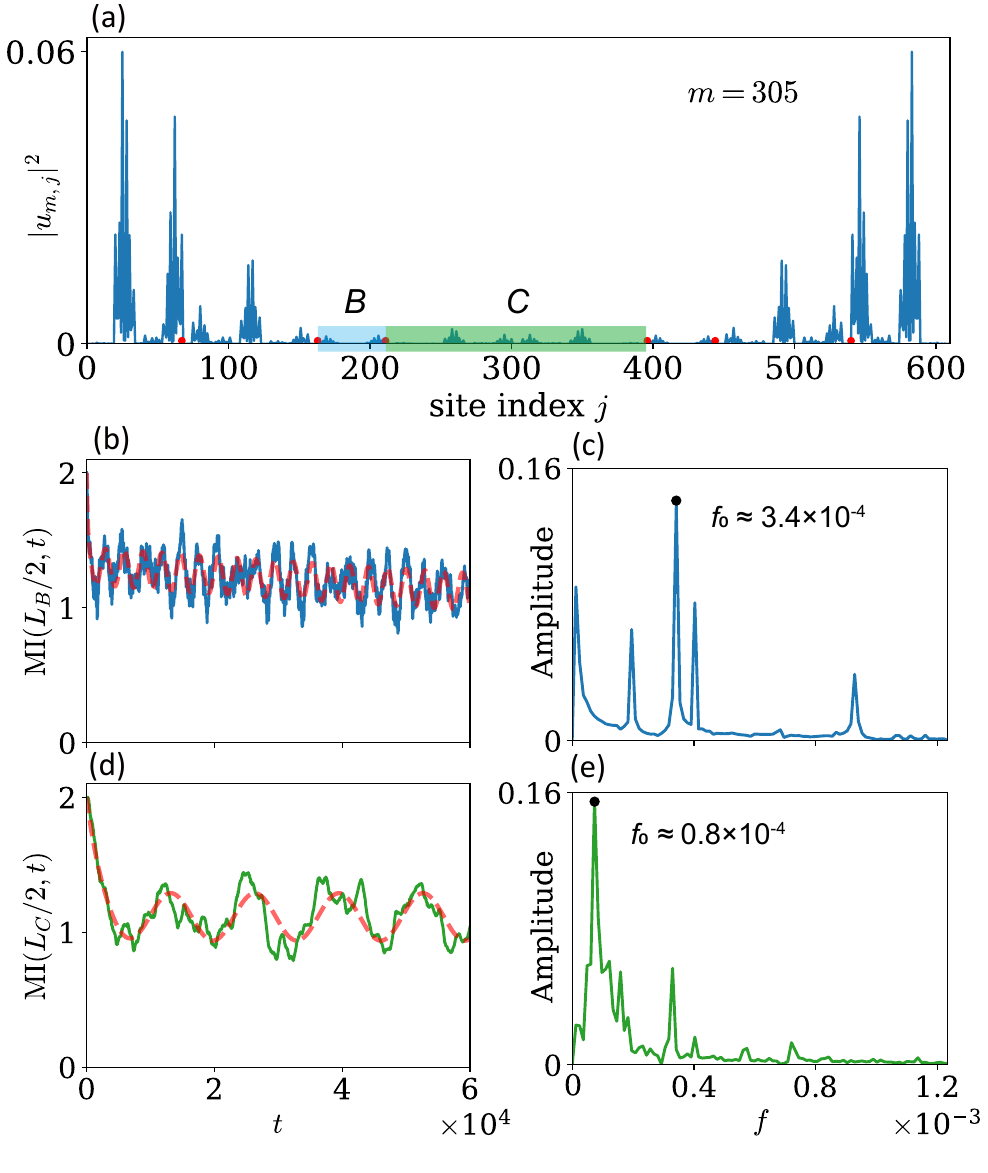}
    \caption{(a) Spatial probability density of the 305th eigenstate in the critical phase with $(\mu, V) = (2.0, 1.0)$ and $L = 610$.
    The density exhibits strong fractal characteristics and is partitioned into several segments by the IDZs (red dots), as indicated by the colored regions $B$ and $C$. 
    {In the pure critical phase of the GAAH model, all eigenstates are critical and exhibit similar fragmentation features separated by IDZs.}
    (b) Time evolution of the SIC within region $B$ ($L_B = 48$) and (c) its Fourier spectrum.
    The initial information is encoded at the center of region $B$, and the subsystem length is set to half of the region. 
    The SIC curve (blue) is {phenomenologically} fitted by $\mathrm{MI}(L_B/2, t) = A_1\exp(-t/\tau_1) + A_2 \exp(-t/\tau_2)+(2 -A_1-A_2)\cos(\omega t)$
    (red dashed), where $\omega = 2\pi f_0$ is fixed by the dominant peak frequency $f_0$ in (c), and $A_1$, $A_2$, $\tau_1$, and $\tau_2$ are free fitting parameters.
    (d)--(e) Corresponding results for region $C$ ($L_C = 185$).
{The two decay times $\tau_1$ and $\tau_2$ characterize distinct information-loss processes: $\tau_1$ describes leakage out of the chosen subsystem of size $L_{\rm region}/2$, while  $\tau_2$ describes the gradual escape out of the entire IDZ-delimited subregion through the near-zero hoppings at its boundaries. The cosine term captures the subregion echoes.}
    }\label{fig3}
\end{figure}

The pronounced spatial inhomogeneity of the critical phase---evidenced by the strong dependence of the SIC 
on the initial qubit position---suggests that the system is internally fragmented. 
A natural mechanism for such fragmentation was proposed in Refs.~\cite{Zhou2023,Zhou2026}, where by employing Avila's global theory~\cite{Avila2015}, 
the authors proved that it is the IDZs in the off‑diagonal hopping terms that partition the system into several nearly decoupled segments.
We refer to the critical phase generated by this mechanism as the IDZ critical phase.
Due to the IDZ fragmentation, wave functions in the IDZ critical phase appear delocalized when inspected globally, yet manifest localized features on local scales. 
This dual character is illustrated in Fig.~\ref{fig3}(a), which displays the density distribution of a critical eigenstate of the GAAH model. The distribution consists of several regions exhibiting strong fractal characteristics, separated by the IDZs (red dots). In the following, we substantiate this picture by exploiting the spatially resolved SIC to directly probe the fragmentation of critical states.

Following the same quench protocol as in the previous study, we investigate how the SIC evolves inside the subregions delimited by the IDZs. We focus on the two regions labeled 
$B$ and $C$ in Fig.~\ref{fig3}(a). For each region, the initial quantum information is encoded at the central site, and the time‑dependent SIC within that region is monitored. 
Figures.~\ref{fig3}(b) and \ref{fig3}(d) show the resulting SIC evolutions, which display pronounced oscillatory behavior. 
To quantify these oscillations, we perform a Fourier spectral analysis, with the corresponding spectra shown in Figs.~\ref{fig3}(c) and \ref{fig3}(e) . 
The spectra exhibit clear dominant peaks, revealing oscillation periods of $T_B\approx 3 \times 10^3$ and $T_C\approx 12 \times 10^3$, respectively. 
Remarkably, the ratio of these periods ($1/4$) closely matches the ratio of the region lengths $L_B/L_C =48/185$, 
indicating that the initial information remains largely trapped within each subregion and undergoes long-lived oscillations (echoes) before eventually escaping.
We refer to these coherent oscillatory revivals, which arise from quasiparticles repeatedly reflected at the subregion boundaries, as {\it subregion echoes}.

The observed subregion echoes find a natural explanation within the quasiparticle picture of quantum quench dynamics, a framework originally developed to describe the ballistic growth and saturation of EE in integrable systems~\cite{Calabrese2005,Alba2017}. In this picture, the initial product state serves as a source of entangled quasiparticle pairs that propagate ballistically with opposite momenta. Within the IDZ‑delimited subregions, these quasiparticles are largely confined by the hopping bottlenecks: instead of spreading across the entire chain, they undergo repeated reflections at the subregion boundaries, producing coherent echoes of the initially encoded information. The resulting oscillation period is expected to scale as $T \propto L_{\text{region}} / \bar{v}$, where $\bar{v}$ is an effective group velocity of the quasiparticles.
Our numerically extracted periods yield a ratio that closely matches the length ratio $T_B / T_C \approx L_B / L_C$, providing strong evidence that the initial quantum information is predominantly carried by quasiparticles trapped within the respective IDZ‑delineated subregions.
As observed in related studies of finite‑size quench dynamics, such oscillatory revivals are a natural consequence of quasiparticles encircling a finite system and returning to their initial location~\cite{Calabrese2005,Modak2020}, here manifested within each effectively decoupled subregion.
The subregion echoes thus provide a dynamical fingerprint of the IDZ-induced bottlenecked connectivity that underpins the multifractal structure of critical states.

We note that not every subregion separated by the IDZs exhibits clear subregion echoes. 
In some subregions, the SIC dynamics show irregular or strongly damped oscillations that do not yield well‑defined dominant peaks in the Fourier spectrum. 
{Even in subregions where subregion echoes exist, the spectra may exhibit additional prominent peaks [Fig.~\ref{fig3}(c)].}
The precise conditions under which subregion echoes emerge remain an open question that warrants further investigation. 
A plausible conjecture is that, although no IDZs are present inside a given subregion, 
certain sites within the subregion may have anomalously small hopping amplitudes due to the incommensurate modulation of the off‑diagonal terms. 
These local hopping bottlenecks can {act as secondary scattering centers that} disrupt ballistic quasiparticle transport, 
preventing the formation of coherent round‑trip reflections and thereby suppressing subregion echoes { or introducing additional frequency components.
In Fig.~\ref{fig3}, regions $B$ and $C$ are chosen as representative examples to illustrate the echo phenomenon and its length scaling, rather than as evidence for its universality across every IDZ-delimited subregion. We emphasize that clear subregion echoes are observed only in sufficiently clean subregions---i.e., those in which the internal hopping profile contains few anomalously small hopping amplitudes.}

\begin{figure}
    \centering
    \includegraphics[width=0.95\linewidth]{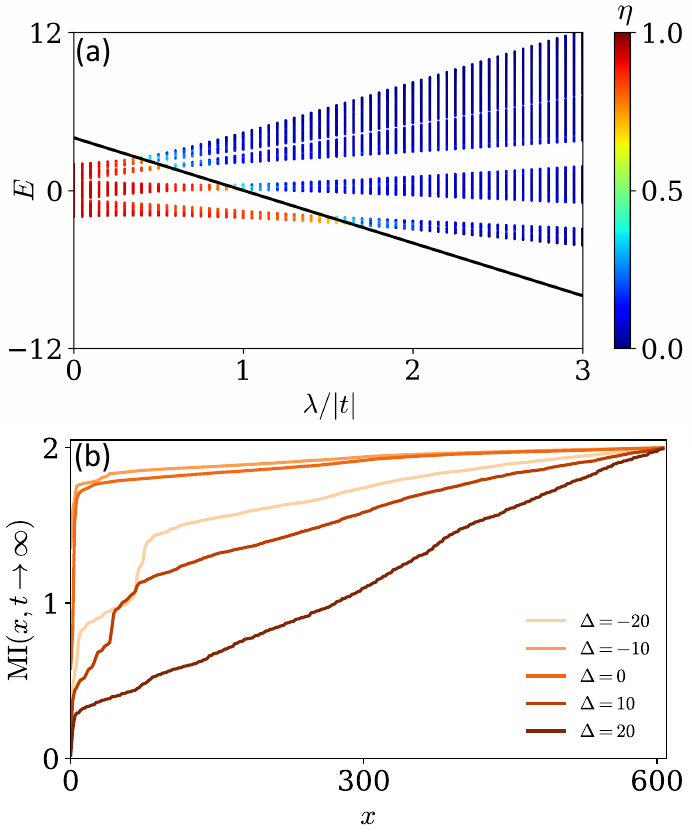}
    \caption{(a) Energy spectrum of the GPDS model for $a = 0.5$ and $L=610$,
    with the $m$th eigenstate colored according to its fractal dimension $\eta$.
    The solid line marks the SPME $E_c$.
    (b) Steady-state SIC in the SPME phase at $\lambda = |t|$
    for different spatial offsets $\Delta$ of the initially entangled site
    $E = L/2 + \Delta$.
    The offset dependence manifests mainly in the extent of the early-stage jump.
    }\label{fig4}
\end{figure}

\section{Comparison with other phases}\label{Sec: Comparison}

The results presented so far establish that the SIC can reveal the IDZ-induced fragmentation of critical states. 
A natural question is whether the SIC signatures identified here are specific to the IDZ mechanism, 
and whether the SIC can distinguish critical phases from other scenarios that also combine extended and localized characteristics. 
To address these questions, we now apply the same analysis to two contrasting cases: a phase with an SPME, 
where extended and localized states coexist in the same spectrum, and a critical phase that does not originate from IDZ fragmentation (dubbed non-IDZ critical phase).
We compare the SIC signatures of the IDZ critical phase with those of two other nontrivial localization scenarios, 
demonstrating both the diagnostic power and the generality of the SIC.

\subsection{SPME phase}\label{Sec: SPME phase}

We first examine an SPME phase with coexisting extended and localized states. 
Although the SPME phase and the critical phase may both exhibit intermediate transport characteristics, 
their underlying mechanisms are fundamentally different, and we show below that the SIC can cleanly distinguish them.

We consider the Ganeshan-Pixley-Das Sarma (GPDS) model~\cite{Ganeshan2015}, described by the
Hamiltonian
\begin{equation}
    H_{\rm GPDS} = -t \sum_{j} \left( c^{\dagger}_j c_{j+1} + { \rm H.c.} \right) + \sum_{j} \epsilon_j c^{\dagger}_j c_j,
\end{equation}
where $t$ is the nearest-neighbor hopping strength, and the quasiperiodic on-site potential is given by
\begin{equation}
    \epsilon_j= 2\lambda \frac{\cos (2\pi\beta j + \phi)}{1 - a\cos (2\pi\beta j+ \phi)},
\end{equation}
with $\lambda$ the potential strength, $\beta = F_{n-1} / F_{n}$, and the system length $L=F_n$. The parameter $a$ distinguishes this model from the
standard AA model; for $a = 0$ the GPDS model reduces to the AA model. We set the global phase to $\phi = 0$.
This GPDS model supports an exact SPME given by the critical energy~\cite{Ganeshan2015}
\begin{equation}
    a E_c = 2\,\mathrm{sgn}(\lambda)\,(|t| - |\lambda|).
\end{equation}
In the parameter region where the SPME intersects the energy spectrum,
eigenstates with energies above $E_c$ are localized, while those below
$E_c$ remain extended.
Figure~\ref{fig4}(a) displays the energy spectrum
of the GPDS model, showing clearly the
SPME phase situated between a pure extended phase and a pure localized phase.

We compute the steady-state SIC following the same quench protocol as in Sec.~\ref{Sec: Steady-state SIC}, starting from the initial
product state $|11\dots100\dots0\rangle$ and quenching to the SPME phase, with the reference
qubit $R$ initially entangled with site $E = L/2 + \Delta$. 
As shown in Fig.~\ref{fig4}(b), the steady-state SIC
exhibits an abrupt initial rise, followed by a steady linear growth, and
finally saturates at $x = L$. The initial rise reflects the information trapping effect of the localized states,
while the subsequent linear growth indicates ballistic transport mediated by the extended states. 
The overall profile of the steady-state SIC in the SPME phase
differs clearly from that in the pure critical phase studied in Sec.~\ref{Sec: Steady-state SIC},
which displays a ramp punctuated by multiple steps. When the initial encoded
site $E$ is varied by the offset $\Delta$, the SIC profile changes mainly in
the extent of the early-stage jump, reflecting the different degree of
localization in the vicinity of the initially entangled site.

\begin{table*}
\centering
\caption{Comparison of SIC signatures across the SPME phase, the IDZ critical phase, and the non‑IDZ critical phase. 
{ In the row labeled ``subregion echoes'', ``yes'' indicates that this signature is allowed or can be observed in principle in that class of models, while ``no'' indicates that such a phenomenon is entirely absent. As discussed in the main text, clear echoes are observed only in sufficiently clean IDZ-delimited subregions.}
}\label{tab:compare}
\begin{tabular}{p{0.28\linewidth}p{0.27\linewidth}p{0.25\linewidth}p{0.16\linewidth}}
\toprule
& \textbf{SPME} & \textbf{IDZ critical} & \textbf{Non‑IDZ critical} \\
\midrule
\textbf{Model}
& GPDS
& GAAH
& SO coupled chain \\
\midrule
\textbf{Steady-state profile}
& Abrupt rise + linear ramp
& Ramp with steps
& Undulating ramp \\
\midrule
\textbf{Initial-site sensitivity}
& Yes (early rise affected)
& Yes 
& Yes \\
\midrule
\textbf{Subregion echoes}
& No
& Yes
& No \\
\bottomrule
\end{tabular}
\end{table*}

\subsection{Critical phase without IDZs}\label{Sec: non-IDZ critical phase}

\begin{figure}
    \centering
    \includegraphics[width=0.95\linewidth]{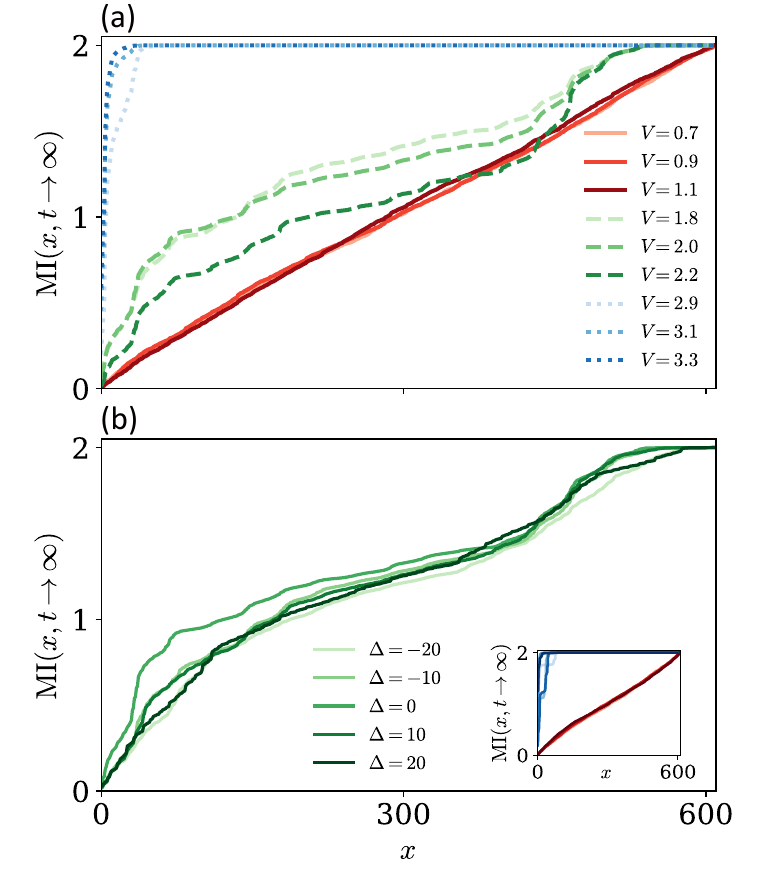}
    \caption{(a) Steady-state SIC after a quantum quench to different phases
    in the SO coupled system for $t_{\rm so} = 0.3$ and $L = 610$.
    Red solid lines: extended phase ($0 < V < 1.4$);
    green dashed lines: critical phase ($1.4 < V < 2.6$);
    blue dotted lines: localized phase ($V > 2.6$).
    The SIC profile exhibits a clear crossover from a ramp-like shape in the
    extended phase to a step-like shape in the localized phase.
    (b) Steady-state SIC for different spatial offsets $\Delta$ of the initially
    entangled site in the critical phase (main panel) and in the extended {and localized phases}
    (inset). In the extended {and localized phases,} only minor variations are observed for
    different offsets, whereas in the critical phase the offsets induce
    pronounced alterations of the SIC curves.}\label{fig5} 
\end{figure}

We next apply the SIC to a non-IDZ critical phase that, in contrast to the GAAH model, is not generated by explicit IDZs in the off-diagonal hopping. 
This example tests whether spatial heterogeneity revealed by the SIC is a universal feature of critical phases or is specific to the IDZ mechanism.

We consider a noninteracting 1D spin-orbit (SO) coupled atomic
chain subjected to a quasiperiodic on-site potential, as introduced in
Ref.~\cite{Wang2020a}. The Hamiltonian reads
\begin{eqnarray}\label{eq: Ham_SO}
    H_{\rm SO} &=& -t_0 \sum_{\langle i,j \rangle}
        \left( c^\dagger_{i,\uparrow} c_{j,\uparrow}
           - c^\dagger_{i,\downarrow} c_{j,\downarrow} \right)
        + \sum_{j} \delta_j \left( n_{j,\uparrow} - n_{j,\downarrow} \right)
        \nonumber \\
      &+& \sum_{j} \left[ t_{\rm so}
        \left( c^\dagger_{j,\uparrow} c_{j+1,\downarrow}
           - c^\dagger_{j,\uparrow} c_{j-1,\downarrow} \right)
        + \mathrm{H.c.} \right].
\end{eqnarray}
Here $t_0$ denotes the spin-conserved nearest-neighbor hopping strength
(set to $t_0 = 1$), {the sum over $\langle i,j \rangle$ runs over nearest-neighbor bonds in both directions,} and $t_{\rm so}$ is the spin-flipped hopping amplitude.
{Note that in the real-space representation, the hopping amplitudes in Eq.~\eqref{eq: Ham_SO} contain no zeros, and therefore cannot directly support the IDZ fragmentation that underlies the critical phase of the GAAH model.}
The quasiperiodic on-site potential
$\delta_j = V \cos (2\pi\beta j + \phi)$
introduces a spin-dependent incommensurate modulation of strength $V$,
with $\beta = F_{n-1} / F_{n}$ and $L=F_n$.
We set the global phase to $\phi = 0$.
For $t_{\rm so} = 0.3$, the model hosts a pure extended phase
($0 < V < 1.4$), a critical phase ($1.4 < V < 2.6$), and a localized phase
($V > 2.6$)~\cite{Wang2020a}.

We compute the steady-state SIC after a quench from the initial state $|\!\uparrow\uparrow\dots\uparrow 00\dots0\rangle$ to each of these phases, 
with the reference qubit $R$ forming a Bell pair with the spin-up state at  site $E = L/2$. 
As shown in Fig.~\ref{fig5}(a), the SIC profile displays a clear crossover from a ramp-like shape in the extended phase to a step-like shape in the localized phase. 
The distinction between the critical and extended phases, however, is less pronounced than that found in the GAAH model: 
the SIC curve in the critical phase exhibits an undulating profile that grows slightly faster than in the extended phase on average, 
but lacks the stepwise ramp characteristic of the IDZ critical phase. 
This is consistent with the fact that, in contrast to the GAAH model, 
this system does not possess explicit IDZs in real space; the critical phase is instead generated by generalized incommensurate zeros in the quasiperiodic on-site potential, as recently proposed for spinful quasiperiodic systems~\cite{Zhou2026}. 
{Although these zeros can be transformed into IDZs in the hopping coefficients of an effective spinless model, the fragmentation mechanism in the original representation is qualitatively distinct from that of the GAAH model.}
When the reference qubit $R$ is instead entangled with the spin-up state at a shifted site $E = L/2 + \Delta$, 
further differences emerge. As shown in Fig.~\ref{fig5}(b), varying $\Delta$ induces visible alterations of the SIC curve in the critical phase, while 
the overall undulating shape persists, indicating spatial heterogeneity within the critical phase. In {the extended  and localized phases}, by contrast, only slight variations are observed (see the inset). 

\subsection{Summary of SIC signatures}\label{Sec: Summary}

The three localization scenarios investigated in this work---the IDZ critical phase, the SPME phase, and the non-IDZ critical phase---exhibit qualitatively distinct SIC signatures, demonstrating the diagnostic power of this probe. Table~\ref{tab:compare} summarizes the key differences.

In the IDZ critical phase, the SIC displays a ramp punctuated by multiple steps in the steady state 
and pronounced subregion echoes arising from quasiparticle confinement within IDZ-delimited subregions.
Both features reflect the internally fragmented structure of critical states. 
The SPME phase, by contrast, produces a steady-state SIC composed of an abrupt initial rise followed by a linear ramp, 
a direct consequence of the coexistence of localized and extended states at different energies. 
No oscillatory dynamics is observed, and varying the initial qubit position alters the extent of the early rise. 
In the non‑IDZ critical phase of the SO coupled model, the steady‑state SIC exhibits an undulating ramp that grows slightly faster than in the extended 
phase on average, but lacks the stepwise structure characteristic of the IDZ critical phase. Spatial heterogeneity nevertheless persists: shifting the 
initial encoding site induces visible alterations in the SIC profile, while its overall undulating shape is preserved, revealing an intrinsic inhomogeneity 
that is not as pronounced as in the IDZ case.
Furthermore, unlike the IDZ critical phase, no regular oscillatory behavior---let alone subregion echoes---is observed in the non‑IDZ critical phase. 
This suggests that the dynamical fingerprints provided by subregion echoes are a distinctive signature of IDZ‑induced fragmentation, 
rather than a generic feature of all critical phases.

Taken together, these results demonstrate that the SIC can {distinguish} IDZ critical phases from both SPME phases and non-IDZ critical phases, 
while also capturing the spatial heterogeneity that appears to be a generic feature of criticality in quasiperiodic systems.

\section{Discussion and Conclusions}\label{Sec: Conclusion}

We have presented a comprehensive study of entanglement and information dynamics in the GAAH model across its extended, critical, and localized phases. By combining the conventional half-chain EE and the spatially resolved SIC, we have demonstrated that entanglement-based probes can faithfully distinguish the three phases and, more importantly, reveal the pronounced spatial heterogeneity underlying the critical states. The SIC uncovers a stepwise ramp in the steady state and subregion echoes in the dynamics. Both features originate from the IDZs in the off-diagonal hopping terms, which partition the system into weakly connected subregions and confine quasiparticle propagation. The quantitative agreement between the observed oscillation periods and the subregion lengths provides { strong} dynamical evidence for the IDZ-induced bottleneck picture. 
By further applying the SIC to two additional scenarios---a phase with an SPME and a critical phase without IDZs---we have demonstrated that the SIC can { distinguish} these cases from the IDZ critical phase through their distinct steady-state profiles, initial-site sensitivities, and the presence or absence of subregion echoes, thereby {demonstrating the diagnostic power of this probe and suggesting its generality across different types of critical phases}.
{We have verified the robustness of these conclusions against finite-size effects (Appendix~\ref{App: finite-size}) and against the choice of initial state (Appendix~\ref{App:  initial-state})}.

Our findings raise several questions that merit further investigation. First, while the quasiparticle picture successfully accounts for the subregion echoes in the non-interacting limit, it remains to be understood how interparticle interactions modify---or possibly wash out---the IDZ-induced fragmentation in interacting generalizations of the GAAH model. Preliminary studies indicate that genuine many-body critical phases can exist in quasiperiodic systems~\cite{Wang2021}, and the SIC may offer a local window into their intricate entanglement and information structures. Second, the interplay between the IDZ mechanism and other forms of quasiperiodic modulation, such as those involving long-range hopping~\cite{Deng2019} or non-Hermitian terms~\cite{Jiang2024,Zhao2025}, could lead to richer fragmentation patterns and merits systematic exploration. Third, extending the SIC analysis to higher dimensions, where critical phases can occupy finite regions of parameter space~\cite{Yang2024}, is an intriguing direction.
Finally, it would also be interesting to investigate how the SIC signatures reported here depend on the particle filling, especially in light of recent predictions of highly nontrivial filling-dependent localization behavior in AA-type models~\cite{Hetenyi2025}.

From the experimental perspective, the SIC is particularly appealing because its measurement requires only a single ancilla qubit and a single forward time evolution. Recent advances in the realization of quasiperiodic potentials in cold-atom systems~\cite{Roati2008,Lucioni2011,Luschen2018,Kohlert2019,An2021,WangY2022,Xiao2021,Shimasaki2024}, photonic lattices~\cite{Gao2025,Zhu2026}, and superconducting circuits~\cite{Li2023,Huang2025} provide a realistic platform for observing the SIC signatures reported here. Specifically, the steady-state SIC spatial profile can be reconstructed by measuring the mutual information between the reference qubit and a variable-sized subsystem, while subregion echoes can be monitored by tracking time evolution of the subsystem entropies~\cite{Chen2025}. We expect that such experiments will not only confirm our predictions but also pave the way for using real-space entanglement and information dynamics as a versatile diagnostic tool for complex quantum phases.

\begin{figure}
    \centering
    \includegraphics[width=0.95\linewidth]{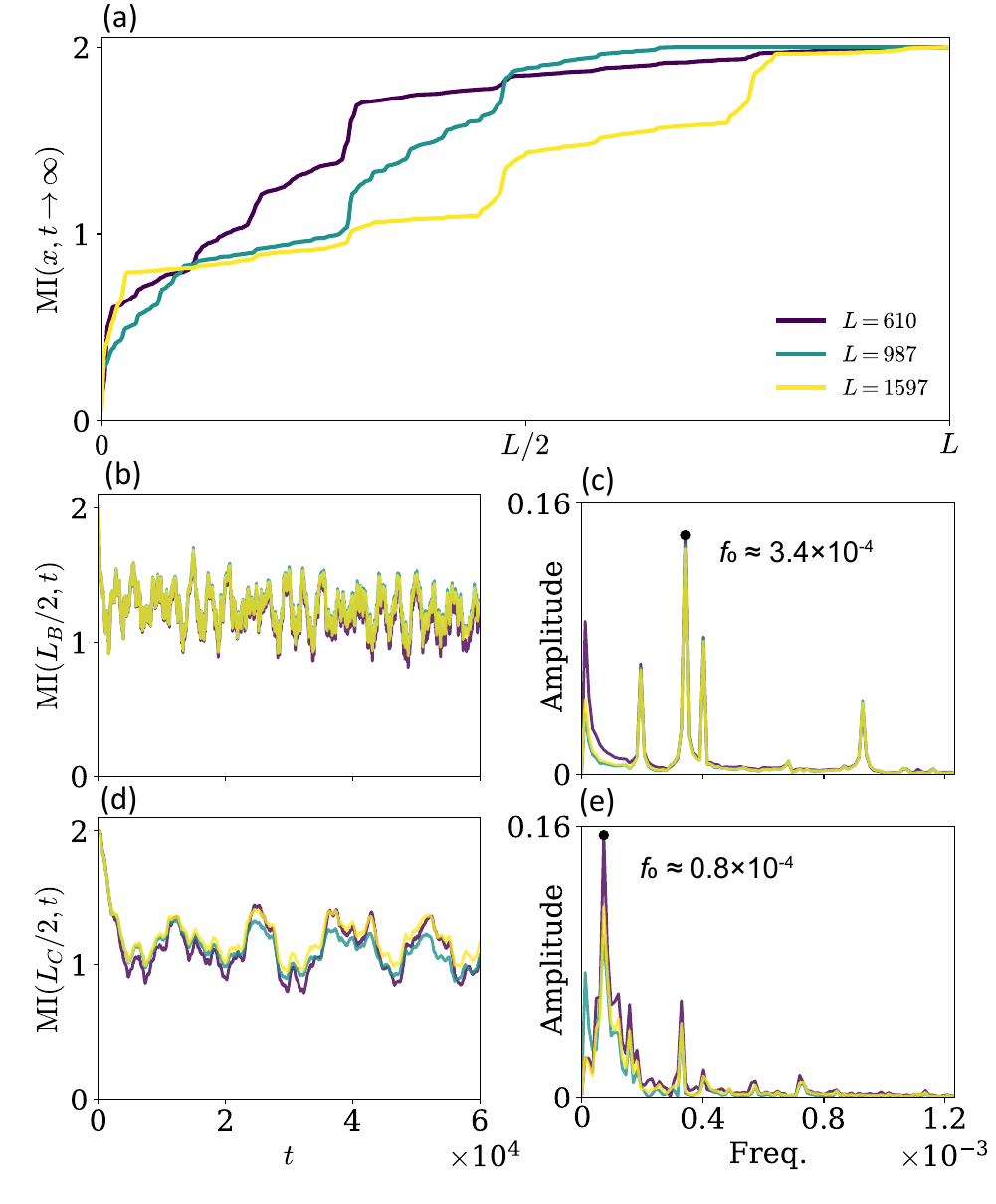}
    \caption{(a) Steady-state SIC for different system sizes $L$ in the critical phase of the GAAH model with parameters $(V,\mu)=(1.0,2.0)$. (b)--(c) Time evolution of the SIC within region $B$ in Fig.~\ref{fig3}(a) and its Fourier spectrum for different $L$. (d)--(e) Corresponding results for region $C$. The encoding and measurement approach is the same as in Sec.~\ref{Sec: SIC Dynamics}.
    }\label{fig6}
\end{figure}

\begin{figure}
    \centering
    \includegraphics[width=0.95\linewidth]{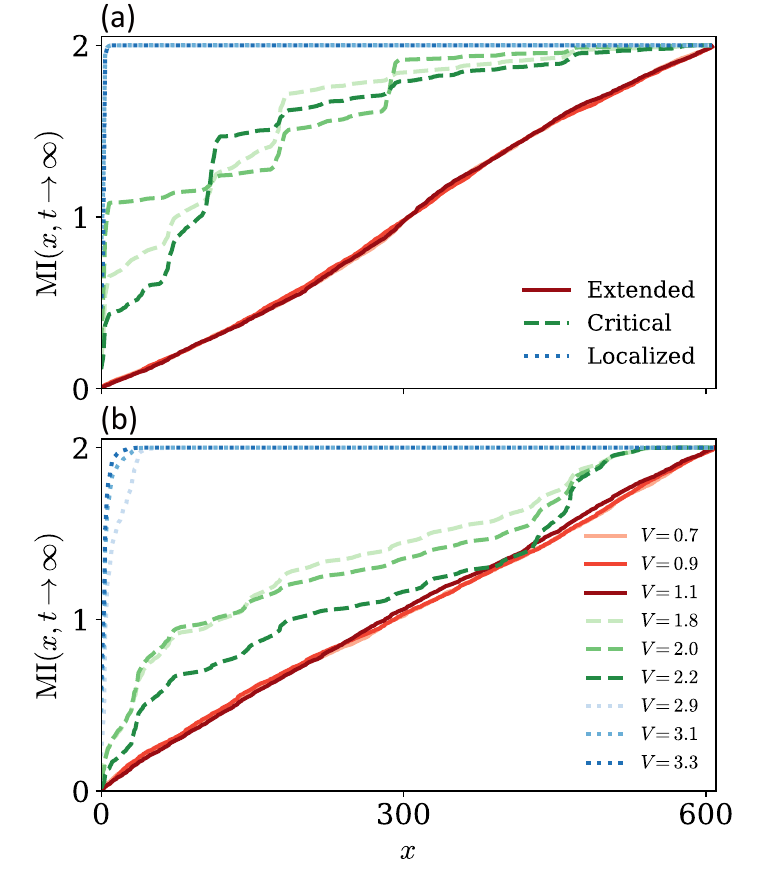}
    \caption{Steady-state SIC from the N\'eel state for different phases of (a) the GAAH model and (b) the SO-coupled model, with the same parameters as in Figs.~\ref{fig2}(a) and~\ref{fig5}(a), respectively, for $L=610$ (and $t_{\rm so}=0.3$ in the SO-coupled case).
    }\label{fig7}
\end{figure}

\section*{Acknowledgements}

We acknowledge insightful discussions with Yucheng Wang and Xin-Chi Zhou.
This work was supported by the National Natural Science Foundation of China (Grants No. 12204187), and
the Innovation Program for Quantum Science and Technology (Grant No. 2021ZD0302000).

\section*{Data Availability}

The data that support the findings of this article are openly available~\cite{Data}.

\appendix

\section{Finite-size analysis of the steady-state SIC and subregion echoes}\label{App: finite-size}

To further substantiate the robustness of our conclusions for the IDZ critical phase of the GAAH model, we repeat the key calculations of the steady-state SIC and the subregion echoes for different system sizes.

Figure~\ref{fig6}(a) displays the steady-state SIC as a function of the subsystem length $x$ for the critical phase at parameters $(V, \mu) = (1.0, 2.0)$ for system sizes $L=F_{14}, F_{15}$, and $F_{16}$. The stepwise structure remains clearly visible across all sizes, suggesting that this feature is robust against finite-size effects.

We next examine the size dependence of the subregion echoes, as shown in Fig.~\ref{fig6}(b). Since the quasiperiodic modulation in Eq.~\eqref{eq1} is determined by the site index and the incommensurate parameter $\beta = F_{n-1}/F_n$, increasing the system size from $L=F_{14}$ to $F_{16}$ changes $\beta$ only slightly, and the resulting shift in the IDZ positions is negligible. Consequently, the IDZ-delimited subregions can be identified at similar real-space locations across different system sizes. We therefore track the subregions corresponding to $B$ and $C$ in Fig.~\ref{fig3}(a) by matching the IDZ positions, and compute the time evolution of the SIC and its Fourier spectrum for each size using the identical encoding and measurement approach as in Sec.~\ref{Sec: SIC Dynamics}. The results show that the SIC dynamics within each subregion remain essentially unchanged across different system sizes, and the subregion echoes persist with their dominant frequencies unaltered. This further confirms that the subregion echoes are intrinsic dynamical signatures of the IDZ-delimited subregions, largely independent of the rest of the system.

\section{Steady-state SIC from the N\'eel state}\label{App: initial-state}

To verify the robustness of the SIC signatures against the choice of initial state, we repeat the steady-state SIC calculations for the GAAH model and the SO-coupled model using the N\'eel state $|101010\dots\rangle$ (or $|\uparrow 0 \uparrow 0 \uparrow 0 \dots\rangle$ for the spinful case) as the initial state. The results are shown in Fig.~\ref{fig7}(a) and (b), respectively.

Comparing with the results obtained from the bipartite product state $|11\dots100\dots0\rangle$ (or $|\!\uparrow\uparrow\dots\uparrow 00\dots0\rangle$ for the spinful case) used in the main text [Figs.~\ref{fig2}(a) and~\ref{fig5}(a)], the steady-state SIC profiles are essentially identical. This demonstrates that the SIC signatures reported in this work are robust against the choice of initial state.

From an experimental perspective, this robustness is practically advantageous: one may choose the initial state that is most convenient to prepare on a given platform without affecting the qualitative SIC signatures. The SIC thus serves as a robust and experimentally adaptable probe of quantum information dynamics in real space.

\bibliography{Critical_SIC}

\end{document}